\documentclass[journal,transmag]{IEEEtran}
\usepackage{graphicx}
\usepackage{xcolor}

\usepackage{amsmath}
\usepackage{cancel}
\usepackage{url}
\usepackage[per-mode=symbol]{siunitx}

\begin{document}
%
\title{Characterizing the Linearity of Magnonic Devices for Radio-Frequency Applications}


\author{\IEEEauthorblockN{Róbert Erdélyi\IEEEauthorrefmark{1}\textsuperscript{,}\IEEEauthorrefmark{2},
Ádám Papp\IEEEauthorrefmark{1}\textsuperscript{,}\IEEEauthorrefmark{2},
Levente Maucha\IEEEauthorrefmark{1}\textsuperscript{,}\IEEEauthorrefmark{2},
Philipp Pirro\IEEEauthorrefmark{3}, 
Matthias Wagner\IEEEauthorrefmark{3}, 
Dieter Ferling\IEEEauthorrefmark{4},
Johannes Greil\IEEEauthorrefmark{5}, \\
Markus Becherer\IEEEauthorrefmark{5}, and
György Csaba\IEEEauthorrefmark{1}\textsuperscript{,}\IEEEauthorrefmark{2}}
\IEEEauthorblockA{\IEEEauthorrefmark{1}Faculty of Information Technology and Bionics, Pázmány Péter Catholic University, Budapest, Hungary}
\IEEEauthorblockA{\IEEEauthorrefmark{2}Jedlik Innovation LLC, Budapest, Hungary}
\IEEEauthorblockA{\IEEEauthorrefmark{3}Fachbereich Physik and Landesforschungszentrum OPTIMAS, Technische Universität Kaiserslautern, Kaiserslautern, Germany}
\IEEEauthorblockA{\IEEEauthorrefmark{4}Nokia Bell Labs, Nokia Corporation, Stuttgart, Germany}
\IEEEauthorblockA{\IEEEauthorrefmark{5}Professorship of Chip-Based Magnetic Sensor Technology, \\ TUM School of Computation, Information and Technology, Technische Universität München}
\thanks{Manuscript received March 2026; revised XXX. 
Corresponding author: Gyorgy Csaba (email: gcsaba@gmail.com).}}


\IEEEtitleabstractindextext{%
\begin{abstract}
Magnonic  devices exhibit strong amplitude-dependent nonlinearities, which are detrimental to  signal integrity in radio-frequency (RF) signal processing applications. They also limit the power that such magnonic devices may process. In this paper we use micromagnetic simulations to characterize the nonlinearity of magnonic RF devices by investigating their intermodulation distortion (specifically third-order intermodulation products, IP$_3$ ). The IP$_3$ is a commonly used metric for RF components in communication systems and allows direct comparison with state-of-the-art electrical counterparts.

\end{abstract}

}

\maketitle

\section{Introduction}
\IEEEPARstart{S}{pin} waves are regarded as promising candidates for serving in beyond-CMOS computational devices and signal processors \cite{Kruglyak2010}. At microwave frequencies, spin-wave  wavelengths may go from millimeters to all the way down  to the nanometer scale making wave-interference-based signal processing and  computing concepts realizable by on-chip devices. They can be excited by frequencies  from the sub-GHz to all the way up to THz ranges, which makes them very well-suited for the next-generation computing and wireless communication systems.


%



Historically, ferrite-based systems have long been  used in microwave processing devices - the reader is referred to \cite{ishak} and the more recent \cite{chumak} for an overview of such devices. The comparably slow propagation speed of and short wavelength of magnons enables compact devices for phase shifting, filtering and power limiting. Chip-scale magnonic devices have recently been used for more for more complex RF processing \cite{Papp2017} \cite{RowlandFIB} or computing \cite{WangInverse}, or even pattern recognition \cite{NatComm} applications.

While there is no 'hard' limit in miniaturizing magnonic devices, most RF applications ideally require the transmission of high powers with high linearity. Magnonic systems, however, exhibit a linear response only at low input signal levels, with nonlinear distortions intensifying rapidly as power density increases. This creates inherent trade-offs between a device’s size and linear power-handling capability. Our study is motivated by the goal of mapping these trade-offs.





In electronic devices, intermodulation (distortion) products (IPs) are commonly used to define the power range when the device nonlinearities remain acceptable. In this work, we leverage this metric to characterize magnonic devices, enabling direct comparison with their electronic counterparts.

Intermodulation  products are closely related to better-known harmonic distortions. Harmonic signals arise from single-tone excitations of a nonlinear system - but they do not necessarily degrade the communication channel since they typically lie well outside the bandwidth of the  channel and can straightforwardly be filtered out. Intermodulation products, arising from the interaction of two tones, are more relevant in device applications, as intermodulation products of close-lying frequencies may produce distortions close to the transmitted  signals, as sketched in Fig. \ref{figure1}. Intermodulation distortion can exert a substantial influence on modulated RF signals utilized for data transmission purposes \cite{Razavi1998}\cite{Wambacq1998}.


To our knowledge, a direct comparison between standard electronic components and chip-scale magonic devices have never been done before - our paper intends to lay the groundwork for assessing potential applications for RF magnonics.

\section{Defining intermodulation products}

If a linear system is excited by a signal at $\omega_1$ frequency, its steady-state output  appears only at this $\omega_1$ frequency.
However, when a sinusoidal input (\begin{math}\omega\end{math}\textsubscript{1}) is introduced to a \emph{nonlinear} system, the resulting output typically displays frequency components that are integer multiples of the input frequency (2\begin{math}\omega\end{math}\textsubscript{1}, 3\begin{math}\omega\end{math}\textsubscript{1}, 4\begin{math}\omega\end{math}\textsubscript{1},...). The component corresponding to the input frequency itself is referred to as the fundamental response, while the additional terms at multiples of the input frequency are termed harmonics. They are exploited in some applications to generate signals at higher frequencies, as in frequency multipliers or harmonic mixers. However, when transmitting modulated signals, as is the case in communication systems, the generation of additional frequencies is undesirable.

\begin{figure}[!t]
\centering
\includegraphics[width=3.5in]{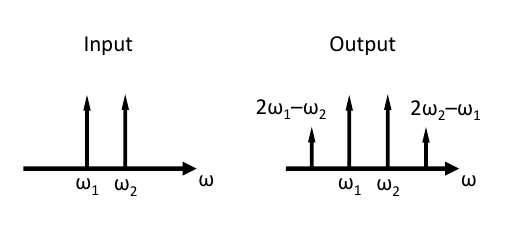}
\caption{Third-order intermodulation products caused by double-tone sine wave stimuli.}
\label{figure1}
\end{figure}

In cases where a nonlinear system is simulatenously subjected to two signals of distinct frequencies (\begin{math}\omega\end{math}\textsubscript{1} and \begin{math}\omega\end{math}\textsubscript{2}), the resulting output may include components that are sums of the integer multiples of the input frequencies. Termed as intermodulation, this phenomenon arises due to the interaction, or "mixing," involving the multiplication of the two input signals. Of particular significance are the third-order intermodulation products at 2\begin{math}\omega\end{math}\textsubscript{1}-\begin{math}\omega\end{math}\textsubscript{2} and 2\begin{math}\omega\end{math}\textsubscript{2}-\begin{math}\omega\end{math}\textsubscript{1}, as shown schematically in Fig. \ref{figure1}. The pivotal observation in this context is that when the difference between \begin{math}\omega\end{math}\textsubscript{1} and \begin{math}\omega\end{math}\textsubscript{2} is small, as it is the case for frequencies of modulated signals, the components at 2\begin{math}\omega\end{math}\textsubscript{1}-\begin{math}\omega\end{math}\textsubscript{2} and 2\begin{math}\omega\end{math}\textsubscript{2}-\begin{math}\omega\end{math}\textsubscript{1} appear in close proximity to \begin{math}\omega\end{math}\textsubscript{1} and \begin{math}\omega\end{math}\textsubscript{2}, possibly falling in the band of interest and hence corrupting the desired component. The situation is similar with higher odd-order intermodulation products, such as fifth or seventh order, but they are usually smaller in magnitude.

Consider a two-tone measurement, where both input tones are applied with an amplitude  $A$. As the value of $A$ rises, the fundamental components at output increase proportionally to $A$, the second-order  harmonics and intermodulation products go with $A^2$ while the third-order intermodulation products surge proportionally to $A^3$ and so on. This is a rather general property of nonlinear systems and directly follows from a Taylor-series expansion of the nonlinear input-output relation \cite{Razavi1998}.

When plotted on a logarithmic scale, the magnitude of the third order  intermodulation products - which is very low at low powers -  demonstrates a threefold steeper growth rate compared to that of the fundamental components and the curves head toward an intersection point, referred to as the third-order intercept point or IP\textsubscript{3}.  

With some assumptions (i.e. that the studied device is a predominantly third-order system \cite{Razavi1998}\cite{Wambacq1998}, the IP\textsubscript{3} point can serve as a single measure of  nonlinearity, facilitating comparisons of linearity among different circuits. 

\section{Methods}

Intermodulation products are most obviously defined on the electrical power flowing through the ports of an electronic device. However, it is also meaningful to interpret the intermodulation products of the power in the magnetic domain. In this case we compare the magnonic power flow in the vicinity of the input and output terminals of the devices. The advantage of this definition is that the results do not depend on the implementation of the input / output transducers and the resulting characteristics describe the magnonic system itself. If input / output waveguide structures are optimized for minimizing insertion loss at the studied frequency band  then only minimal differences are expected between the IP\textsubscript{3} values defined through the ports \cite{wg_model} \cite{Connelly2021}.

Intermodulation products are examined through micromagnetic simulations. We study a Yttrium Iron Garnet (YIG) waveguide  (width 10 $\mathrm{\mu m}$,  film thickness 500 nm and distance between the transducers is  100 $\mathrm{\mu m}$)  in forward volume geometry, using the well-established MuMax3 code \cite{Vansteenkiste2014}.  We always used $\omega_1=2\pi\times 2 \textrm{ GHz}$ and $\omega_2=2\pi\times 2.1 \textrm{ GHz}$. So the intermodulation products appear at $1.9\textrm{ GHz}$ and $2.2\textrm{ GHz}$ frequencies. Futher details of the simulations are given in the Appendix.  

\begin{figure}[!t]
\centering
\includegraphics[width=3.5in]{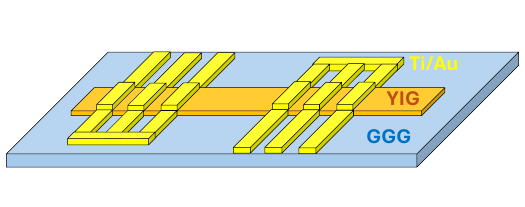}
\caption{Schematic of a standard two-port magnonic system composed of a ferromagnetic thin film, an input antenna, and an output antenna.}
\label{figure2}
\end{figure}

We numerically determined the magnetostatic energy density close to the excitation waveguide and at $l=15$ $\mathrm{\mu m}$ distance from it, where the output waveguide is assumed (see Fig. \ref{figure2}). This $E_\mathrm{dens}(f)$ energy density was determined separately for each frequency of interest.

Our  calculation is based on  finding the $E_\textrm{dens}(f)$ magnetostatic energy density in the magnetic film as the function of the $f$ frequency. $E_\textrm{dens}(f)$ is determined from the steady-state ${\bf M}(f)$  magnetization amplitude and the steady-state ${\bf B}(f)$ flux density, using the formula $E_\mathrm{dens}(f)= -0.5\mathbf{B_{demag}}(f)\mathbf{M}(f)$. The $\mathbf{B_{demag}}(f)$ and $\mathbf{M}(f)$ components are numerically determined by extracting the magnetizaton and field amplitude from the time-domain simulation data  at each $f$ frequency of interest  (see Appendix for details). Since we work in the dipole-dominated regime, exchange energies are not considered even if they could have been straightforwardly take into account by their effective field.

Knowing the $E_\textrm{dens}(f)$ energy density for each propagating mode, the $S$ cross section of the YIG  and the $v_g(f)$ group velocity, the  time-averaged power  flow through the YIG cross section can be calculated by: $P_{SW}(f)= S E_\textrm{dens}(f) v_g(f) $.

The curves that yield to the IP\textsubscript{3} are calculated from $P_{SW}(f)$ curves at the output at the $\omega_1$ and $\omega_2$ fundamental and $2\omega_1-\omega_2$ and $2\omega_2-\omega_1$ frequencies. The output powers are plotted as the function of incoming spin-wave power.


\section{Results and discussion}

\subsection{Quantifying the linearity of magnonic conduits}

\begin{figure}[!t]
\centering
\includegraphics[width=3.5in]{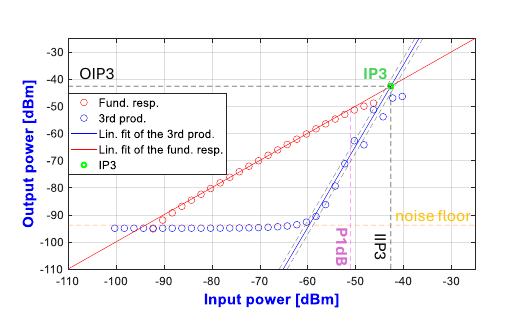}
\caption{Fundamental response and third order intermodulation product 15 $\mathrm{\mu m}$ far from the microstrip antenna in forward volume mode. Spin waves are excited in a 500 nm thick YIG layer.}
\label{figure5}
\end{figure}


We can extract the IP\textsubscript{3} from two simulations that use single and two-tone excitations. Fig. \ref{figure5} illustrates the fundamental response (at \begin{math}\omega\end{math}\textsubscript{1}) and one of the third-order intermodulation products (at 2\begin{math}\omega\end{math}\textsubscript{2}-\begin{math}\omega\end{math}\textsubscript{1}) as a function of input power. The YIG magnonic conduit is  excited in forward volume (FV) mode by the field of a waveguide.  As we progressively increase the input signal amplitude in the single-tone case at \begin{math}\omega\end{math}\textsubscript{1}, the output signal at \begin{math}\omega\end{math}\textsubscript{1} (red circles) displays a linear growth in response to input power. Beyond a certain applied input power, the output power starts to  gradually decrease with respect to the input power. The difference reaches 1 dBm at the magenta dashed line in Fig. \ref{figure5}. This point is commonly referred to as the $IP_1$ point. At this point the device goes into compression and becomes non-linear, producing distortion, harmonics and intermodulation products. The blue circles show the response at 2\begin{math}\omega\end{math}\textsubscript{2}-\begin{math}\omega\end{math}\textsubscript{1}. At lower input power levels, this response is hardly detectable, but increases steeply.  The linear segment is then followed by a breakdown in trend. IP\textsubscript{3} is found  as the input power at which the fitted lines of the linear parts of the responses at 2\begin{math}\omega\end{math}\textsubscript{2}-\begin{math}\omega\end{math}\textsubscript{1} and the fundamental frequency intersect. The slopes of the fitted lines for these responses should be 3 and 1, respectively. This criterion is inherently satisfied for the fundamental response. For the third-order intermodulation product, a linear fit with a fixed slope of 3 was applied. The confidence band for this fit is represented by dashed gray lines surrounding the fitted line. IP3 is depicted by a green circle.

Figure \ref{figure5} also indicates a 'noise floor' (yellow line). This noise level is calculated assuming a $\Delta f$=100 MHz bandwidth at room temperature (see Appendix for details). The  $\Delta f$=100 MHz bandwidth is an arbitrary choice, but this makes a good fit to the lowest power level detectable at the fundamental and intermodulated frequencies. This noise is a combination of 'real' magnetic noise, coming from the stochastic  Landau-Lifshitz equation and errors such as numerical accuracy and finite simulation time.

According to Fig. \ref{figure5} the P1dB point of the YIG-based magnonic device is approximately -51 dBm, and the IP3 point is around -43 dBm. The simulated 8 dB difference between the IP1 point and the IP3 is reasonably close to the (approximately) 9.6 dB difference expected from the theory of a third-order system \cite{Razavi1998} and confirms the dominant influence of third-order effects. 

The -51 dBm value at the P1dB point corresponds to $P_\textrm{dens}=1.5  \mathrm{nW/\mu m^2}$ power density over the cross section of the YIG film.

\subsection{Scaling rules and comparison with electronic counterparts}

To put the above numbers in context, we  compare magnonic devices  to their electronic counterparts. Perhaps the easiest comparison is made to RF phase snifters, which are similarly two-port, passive devices \cite{Lee2019}, \cite{Ma2013} \cite{Ma2014}. 

The total power handling capacity of the magnonic conduit is estimated as P1dB point. The -51 dBm value we calculated may be scaled up by  making the device wider - scaling the width by a factor of 100 (from 10 $\mathrm{\mu m}$ to 1 mm) increases P1dB to -31 dBm. 


Still, the power carrying capacity of the magnonic conduit is significantly lower than that of commercial phase shifters for general use, which are typically operated in the +20 dBm range. However, in special designs, the order of the components in a transceiver circuit can be optimized, as shown in \cite{kebe} for phase shifters in phased array systems. Accordingly, magnon devices can be used very effectively in radio front ends before a power amplifier in the transmit section or after the low-noise amplifier in the receive section, where signals typically occur below -31 dBm. 

In a similar fashion, the IP3 point may be scaled up to -23 dBm by increasing device width  by a factor of 100 - this compares to the +25 dBm to 45 dBm value in typical commercial components for general use – and allows for usage in dedicated applications.

The dynamic range (DR) is the difference between the smallest above-noise signal and the P1dB point. In our case this is $ \mathrm{DR}=52 \mathrm{dB}$. This is lower than the DR in most of the above-cited examples, but acceptable for several applications.

\section{Conclusion}

The strong nonlinearity of spin waves is attractive for nonlinear computational tasks, but it can be a challenge in linear signal-processing applications. Nonlinear distortion limits the dynamic range of spin-wave devices, and nonlinear effects occur at relatively low powers for spin waves. We demonstrated a computational method to quantify the nonlinearity of magnonic devices, based on estimation of the third-order intersect point ($\mathrm{IP_3}$). The $\mathrm{IP_3}$ product is  a widely used measure and it have been used to characterize for magnonic RF components in the past (\cite{ref:ip31}, \cite{ref:ip32}). But we are not aware of prior work on determining $\mathrm{IP_3}$ point for micron-scale devices and independently of the antenna transducer geometry. 

We exemplified the methodology on one particular setup, with a somewhat arbitrarily-chosen geometry and frequency. No attempt was made to optimize the device characteristics. The P1dB and $\mathrm{IP_3}$ values may be slightly different for other geometries or propagation modes, and both will likely grow toward higher frequencies - Hamiltionian models suggest a linear increase of the P1dB point with $\omega$ \cite{ref:krivosik}. Increasing the film thick thickness (and its cross section) will proportionally increase the power flow.  

The estimated P1dB and IP3 values help identify suitable applications for micron-scale magnonic devices. Although their dynamic range and power-handling capability are smaller than those of typical electronic components, their figures of merit remain well within the useful range for many RF applications.

We hope that this methodology helps defining application drivers for emerging magnonic devices.

\appendix[Details of micromagnetic simulations]
\label{sec:appendix}

 Micromagnetic simulations were conducted using MuMax3 \cite{Vansteenkiste2014}, an open-source, GPU-accelerated micromagnetic simulation software. MuMax3 is widely used in the research community and has been extensively tested and shows great predictive power in problems where micromagnetic description is appropriate. We used periodic boundary conditions in the micromagnetic simulations to ensure that a perfect plane wave is formed at the excitation antenna and  the power scales linearly wdth the transucer width. A Forward Volume (FV) geometry was considered, that is, when the magnetic field is applied out of plane and spin waves propagate isotropically in the $xy$ film plane. 

All simulations were performed using the temperature-dependent module of MuMax, using $T=300$ K. Besides adding noise to the simulations, temperature slightly affects the magnitude of the IP3 products.

 We used the well-known Kalinikos-Slavin  formula  \cite{kalinikos_slavin} to determine the magnonic wavelength at $\omega=2\pi\times 2$ GHz frequency at a any given bias field. We used  a $B_z = 235.9 \textrm{ mT}$ external bias field that gives a wavelength of $10 \mu$m - this wavelength is excited efficiently by the input transducers.  
 
 To generate spin waves we have chosen a coplanar waveguide with a central conductor   and used   FEMM~\cite{FEMM}, an open-source 2D finite element analysis software package to calculate the magnetic field distribution around the waveguide, assuming a uniform current distribution. We scaled the currrent density up / down to achieve different power levels in the magnonic conduit. The waveguide was designed to efficiently excite $\lambda=10\mu $m wavelength spin waves.

While the simulations used realistic waveguide geometries, our results remain independent of the waveguide since we study power flow in the magnonic domains. Using physically realizable waveguides in the simulations is still beneficial, as it prevents artifacts (such as spurious modes) in the propagation wavefront.


The micromagnetic simulation gives $\mathbf{M}(t)$ and $\mathbf{B}(t)$ at every point of the simulation region. The propagating power at different frequencies was extracted from the $\mathbf{M(\omega)}$ and $\mathbf{B(\omega)}$ complex apmplitudes, i.e. the magnetization and flux density oscillation ampitudes at the frequency components of interest.  The $\mathbf{M(\omega})$ and $\mathbf{B(\omega})$  values can be extracted by a multiplication of a sine and cosine function at the given frequency. Simply speaking it corresponds to a 'lock in' measurement on the simulation data.

We estimated the noise floor to be :

\begin{equation}
P_{thermal} = 10 \log_{10} \left( \frac{k_B T \Delta f}{1 \ \text{mW}} \right)
\label{eq:noise}
\end{equation}

where
\begin{align*}
k_B & : \text{Boltzmann constant},\\
\Delta f & : \text{frequency bandwidth in the simulations: 100 MHz},\\ 
T & : \text{temperature: 300 K}.\\ 
\end{align*}

The power levels calculated this way coincide with the lowest power levels we could extract from numerical simulations.


\section*{Acknowledgment}

The authors acknowledge support from the European Union within the HORIZON-CL4-2021-DIGITAL-EMERGING-01 Grant No. 101070536 MandMEMS. A.P. received support through the Bolyai János Research Fellowship of the Hungarian Academy of Sciences.

\ifCLASSOPTIONcaptionsoff
  \newpage
\fi

%








\end{document}